\RequirePackage{fixltx2e}
\documentclass[english,twocolumn]{article}
\usepackage[T1]{fontenc}
\usepackage[latin9]{inputenc}
\usepackage{graphicx}

\makeatletter
\let\oldmaketitle\maketitle
\let\maketitle\relax
\usepackage{natbib}
\setcitestyle{aysep={}} 
\setcitestyle{authoryear,open={},close={}}

\makeatother

\usepackage{babel}
\begin{document}

\title{Understanding In-line Probing Experiments by Modeling Cleavage of
Non-reactive RNA Nucleotides}

\author{Vojt\v{e}ch Ml\'{y}nsk\'{y} and Giovanni Bussi \\ \\
vmlynsky@sissa.it, bussi@sissa.it \\
Scuola Internazionale Superiore di Studi Avanzati di Trieste,\\
via Bonomea 265, 34136, Trieste, Italy}

\maketitle
\twocolumn[
\begin{@twocolumnfalse}
\oldmaketitle
\begin{abstract}
Ribonucleic acid (RNA) is involved in many regulatory and catalytic processes in the cell. The function of any RNA molecule is intimately related with its structure. In-line probing experiments provide valuable structural datasets for a variety of RNAs and are used to characterize conformational changes in riboswitches. However, the structural determinants that lead to differential reactivities in unpaired nucleotides have not been investigated yet. In this work we used a combination of theoretical approaches, i.e., classical molecular dynamics simulations, multiscale quantum mechanical/molecular mechanical calculations, and enhanced sampling techniques in order to compute and interpret the differential reactivity of individual residues in several RNA motifs including members of the most important GNRA and UNCG tetraloop families. Simulations on the multi ns timescale are required to converge the related free-energy landscapes. The results for uGAAAg and cUUCGg tetraloops and double helices are compared with available data from in-line probing experiments and show that the introduced technique is able to distinguish between nucleotides of the uGAAAg tetraloop based on their structural predispositions towards phosphodiester backbone cleavage. For the cUUCGg tetraloop, more advanced ab initio calculations would be required. This study is the first attempt to computationally classify chemical probing experiments and paves the way for an identification of tertiary structures based on the measured reactivity of non-reactive nucleotides.
\end{abstract}
\end{@twocolumnfalse}
] 

\section{Introduction}

Ribonucleic acid (RNA) participates in several kinds of cellular processes,
which involve the transmission of genetic information, the synthesis
of proteins, the cellular differentiation and development, the regulation
of gene expression and enzyme-like catalysis (\cite{Bevilacqua_NAtChem2012,Lee_lncRNAs_Genetics2013,Miska_sRNAs_NatRevMolCellBiol2014}).
Detailed information about RNA secondary structures is a preliminary
step required for tertiary structure determination and, ultimately,
for understanding RNA function (\cite{Walter_Methods2009}). Identification
of specific small RNA motifs, like RNA tetraloops, is particularly
important as they stabilize larger RNA structures and can be involved
in RNA-RNA or RNA-protein interactions (\cite{Hall_tetraloops_RNA2015}).
Indirect information about RNA secondary structure is usually obtained
by chemical probing experiments (\cite{Culver_Methods-Enzymology_2009,Weeks_Curr-Opi-Struc-Biol_2010,Spitale_NatChemBiol2015}).
Among those, selective 2'-hydroxyl (2'-OH) acylation characterized
by primer extension (SHAPE, \cite{Merino_SHAPE_JACS2005}) and in-line
probing (\cite{Soukup-Breaker_RNA1999}) experiments provide sequence
independent structural information on RNA at single-nucleotide resolution
(\cite{Weeks_Curr-Opi-Struc-Biol_2010}).

In-line probing characterizes backbone mobility by structural dependent
phosphodiester cleavage, which breaks RNA molecules into segments
at distinct positions (\cite{Soukup-Breaker_RNA1999}). Unpaired nucleotides
within single stranded RNA regions are often unstable and degrade
over time (\cite{Reynolds_NAR1996,Welch_Biochemistry1997}). The chemical
reaction, termed as an internal trans-esterification, starts with
2'-OH attack of neighboring phosphate moiety and results with 2',3'-cyclic
phosphate and 5'-hydroxyl termini (\cite{Soukup-Breaker_RNA1999,Lilley_Trends2003}).
The same RNA backbone cleavage (or RNA degradation) is performed by
self-catalytic systems called RNA enzymes (ribozymes, \cite{Doudna-Cech_ribozymes_Nature2002,Scott_Ribozymes_CurrOpiStrucBiol2007,Lilley_Book2008})
and by ribonuclease A (RNase A, \cite{Raines_ChemRev1998}), although
in these cases with significantly higher rate constants. By comparing
nucleotides from various RNA motifs, Soukup and Breaker observed a
relation between the cleavage rate constant and the in-line attack
angle of the scissile phosphate, i.e., the angle between O2' oxygen,
the adjacent phosphorus, and O5' oxygen (\cite{Soukup-Breaker_RNA1999}).
Later, they defined the ability to bring the active site towards the
in-line attack conformation (the in-line attack angle close to 180$^{\circ}$)
as one of catalytic strategies for the phosphodiester backbone cleavage
used by ribozymes (\cite{Breaker-et-al_RNA2003,Emilsson_RNA2003}).
Since then, in-line probing is routinely used in studies of riboswitches,
where the binding of a small molecule (ligand) results in a conformational
change of the whole RNA molecule (\cite{Breaker_NatRev2004,Breaker_MethodsMolBiol2008,Batey_AnnuRevBiophys2008,Batey_PerspBiol2011}).
In general, chemical probing experiments are typically employed to
identify unpaired and flexible nucleotides, allowing one to choose
among different predicted secondary structures. However, it must be
noticed that not all the unpaired nucleotides are usually reactive.
The reactivity pattern of unpaired nucleotide could in principle provide
a wealth of information that is usually discarded. To the best of
our knowledge, the pattern of reactivity of specific motifs have never
been analyzed in detail.

Computational techniques are an established tool for the investigation
of structural and dynamical properties of RNA at an atomistic level
(\cite{Schlick_Book2010,Cheatham-Case_Biopolymers2013,Sponer-Methods2013})
and could in principle allow for an investigation and interpretation
of reactivity patterns in RNA. In particular, quantum mechanical/molecular
mechanical (QM/MM, \cite{Warshel_qmmm_JMB1976}) approaches, where
only the reactive portion of the system is described at the QM level,
have been used to characterize cleavage reactions within catalytic
RNA systems (\cite{Banas_JPCB2008,Nam_JACS2008,Nam_RNA2008,York-hammerhead_JACS2008,Mlynsky_JPCB2011,Hummer_qmmm-RNaseA_JACS2011,Xu_qmmm-ribosome_JACS2012,York-Harris_PNAS2013,Hammes-Schiffer_HDVr_JACS2014,Mlynsky_PCCP2015,Dubecky_Biopolymers2015,Hammes-Schiffer_glmS_JACS2015,Bevilacqua_HDVr_JACS2015,Casalino_JACS2016}).
Single point QM/MM calculations evaluate potential energy surfaces
neglecting entropic contributions. The reconstruction of free-energy
surfaces (FES) along the reaction pathway requires combination of
QM/MM calculations with molecular dynamics (MD) simulations and enhanced
sampling techniques (\cite{palermo2015catalytic}). In this context,
semiempirical (SE) methods allow for a reasonable compromise between
accuracy and efficiency (\cite{Christensen_Kubar_ChemRev2016}), allowing
statistically converged FES to be computed. Two SE methods have been
carefully parametrized for reactions involving the phosphate moiety
(\cite{Nam_AM1d_JCC2007,Yang_DFTBPR_JCTC2008}).

In this paper we combine QM/MM calculations and enhanced sampling
methods to model phosphodiester backbone cleavage of nucleotides from
three model systems: one tetraloop from each of the GNRA and UNCG
family (R = purine and N = any nucleotide) and a double stranded RNA
(dsRNA). Regions undergoing the cleavage reaction were described by
a DFTB3 SE method, which allowed us to perform simulations on tens-of-ns
timescales and obtain converged free-energy landscapes. Our calculations
required to design a putative reaction pathway with a number of restraints
in order to overcome persisting shortcomings within parameterization
of the DFTB3. However, we were able to differentiate among nucleotides
within simple motifs by comparing their activation free-energy barriers.
Computational results were validated against available experimental
data from in-line probing measurements. To our knowledge, this represents
the first attempt to design computations in order to understand and
mimic in-line probing experiments and, more generally, to investigate
the phosphodiester backbone cleavage within non-reactive RNA nucleotides,
i.e., not considering the active centers of RNA catalytic motifs.

\section{Results}

We performed combined QM/MM-MetaD simulations (see \emph{Materials
and Methods}) and reconstructed FES relative to the phosphodiester
cleavage reaction for nucleotides within three simple RNA motifs:
uGAAAg tetraloop, dsRNA (GC-duplex) and cUUCGg tetraloop. For the
tetraloops we computed the reactivity for the unpaired residues and
for the closing base pairs. For the duplex we chose 3 consecutive
non-terminal residues from each strand (Figure S1 in the Supporting
Information (SI)). We thus analyzed relative differences in reactivities
among 6 nucleotides for each system by comparing $\Delta G^{\ddagger}$
cleavage barriers along reaction pathway designed to be equivalent
for all nucleotides from different RNA motifs. Since our intention
was not to give insight into the mechanism of phoshodiester backbone
cleavage, possible involvement of other RNA groups, water molecules,
and ions was omitted and any mechanistic interpretation deliberately
neglected.

\begin{figure*}
\begin{centering}
\includegraphics[width=0.55\textwidth]{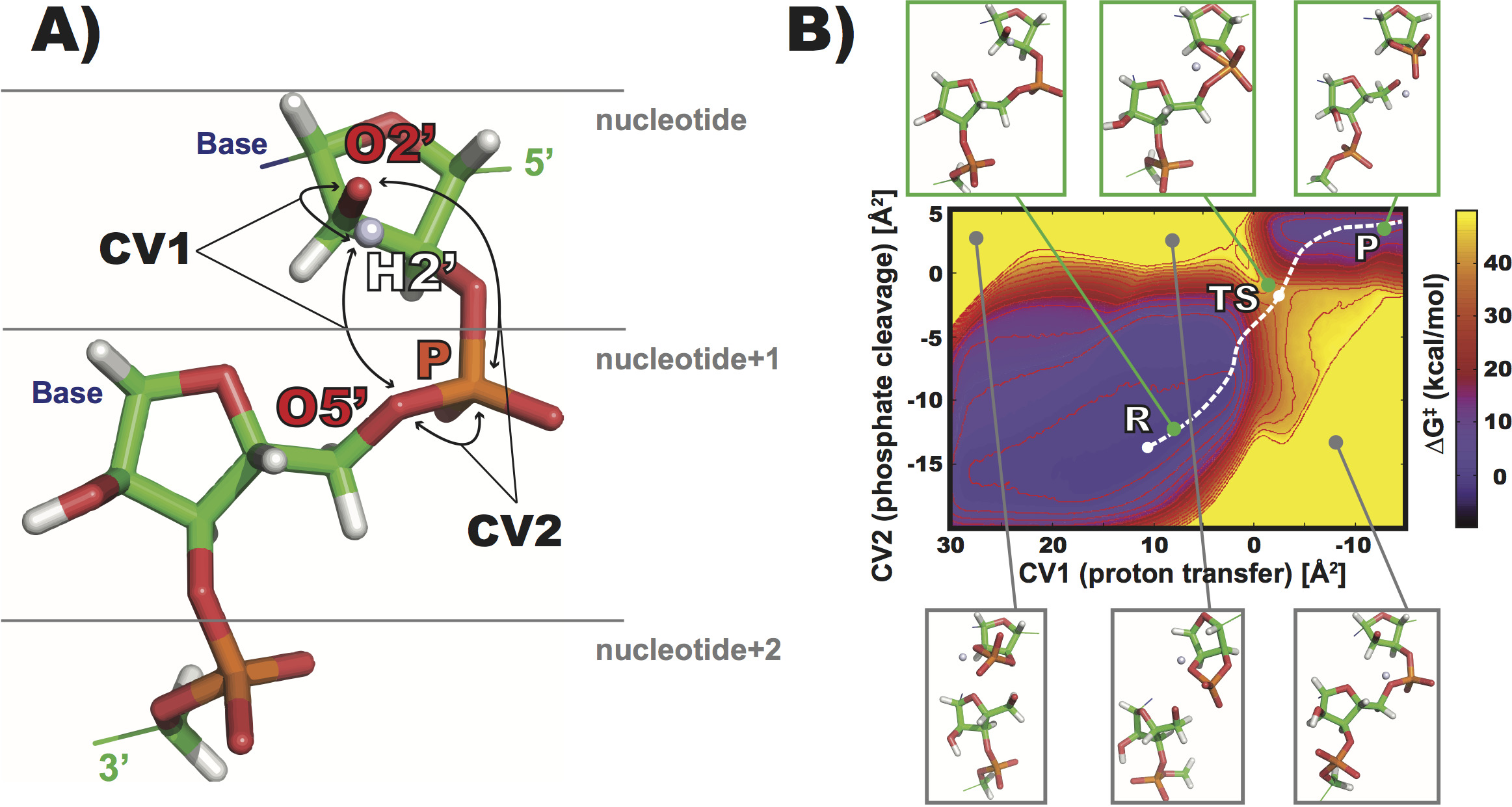}
\par\end{centering}
\caption{\label{fig1} (A) Selection of QM region and definition of collective
variables (CV's). QM region for particular nucleotide contained ribose
ring (reactive 2'-OH group), phosphate with ribose ring of nucleotide+1,
and phosphate moiety of nucleotide+2. Bases were described at MM level
of theory (AMBER \emph{ff}14). FES of phosphate cleavage reaction
was defined by two CV's. CV1 described the proton transfer (difference
of squared (O5'...H2') and (O2'...H2') distances) and CV2 described
the phosphate cleavage (difference of squared (O5'...P) and (O2'...P)
distances). (B) FES of phosphodiester backbone cleavage for G4 nucleotide
from uGAAAg tetraloop. Snapshots in boxes show different conformation
and reaction states of QM atoms defined by specific values of CVs.
The proton from 2'-OH is represented as unbound to be clearly visible.
Green boxes show geometries close to R, TS and P states, whereas grey
boxes display structures energetically penalized by restraints (see
SI for details). The white dashed line marks the estimated reaction
coordinate.}
\end{figure*}

The computed FES profiles mapped $\Delta G$ changes along two tracked
distance-based collective variables (CVs, Figure \ref{fig1}A) to
describe the proton transfer and the phosphodiester cleavage. The
most relevant information that we want to extract is the $\Delta G^{\ddagger}$
cleavage barrier, which was estimated as explained in \emph{Materials
and Methods}. The FES displays two energy minima containing reactant
(R) and product (P) state geometries (Figure \ref{fig1}B). The R
state minimum is very broad because the simulation is allowed to sample
all the possible geometries including different orientations of the
active 2'-OH group. This is necessary for the accurate estimation
of $\Delta G^{\ddagger}$ barriers. On the contrary, we restrained
the extensive separation of the RNA molecule after the cleavage reaction,
leading to a narrower free-energy minimum associated to P state. The
complete exploration of P state geometries would make convergence
very difficult and is irrelevant for a proper estimation of the cleavage
barrier.

The FES profiles for nucleotides within three different RNA motifs
(uGAAAg and cUUCGg tetraloops, GC-duplex) are qualitatively very similar
among each other (see Figure S2 in SI for all computed FES). The R
and transition states (TS) have slightly different locations for particular
nucleotides within each RNA motif, but no relevant correlations. For
instance, purine/pyrimidine or tetraloop/duplex differences, were
not detected. The computed $\Delta G^{\ddagger}$ barriers after 40
ns-long QM/MM-MetaD simulations were in the range between 40.0 and
44.5 kcal/mol. Note that the reaction coordinates are affected by
the applied restraints and approximations, which forced the proton
from the 2'-OH to be kept around the direct pathway (Figure \ref{fig1}).
Furthermore, the reported cleavage barriers are expected to be overestimated
due to the complete exclusion of scenarios involving proton transfer
through nonbridging oxygens (nbO) of the adjacent phosphate and/or
through other proximal RNA groups. However, this allows for a consistent
estimation of $\Delta G^{\ddagger}$ values and their comparison across
nucleotides within different RNA motifs.

\begin{figure*}
\begin{centering}
\includegraphics[width=0.75\textwidth]{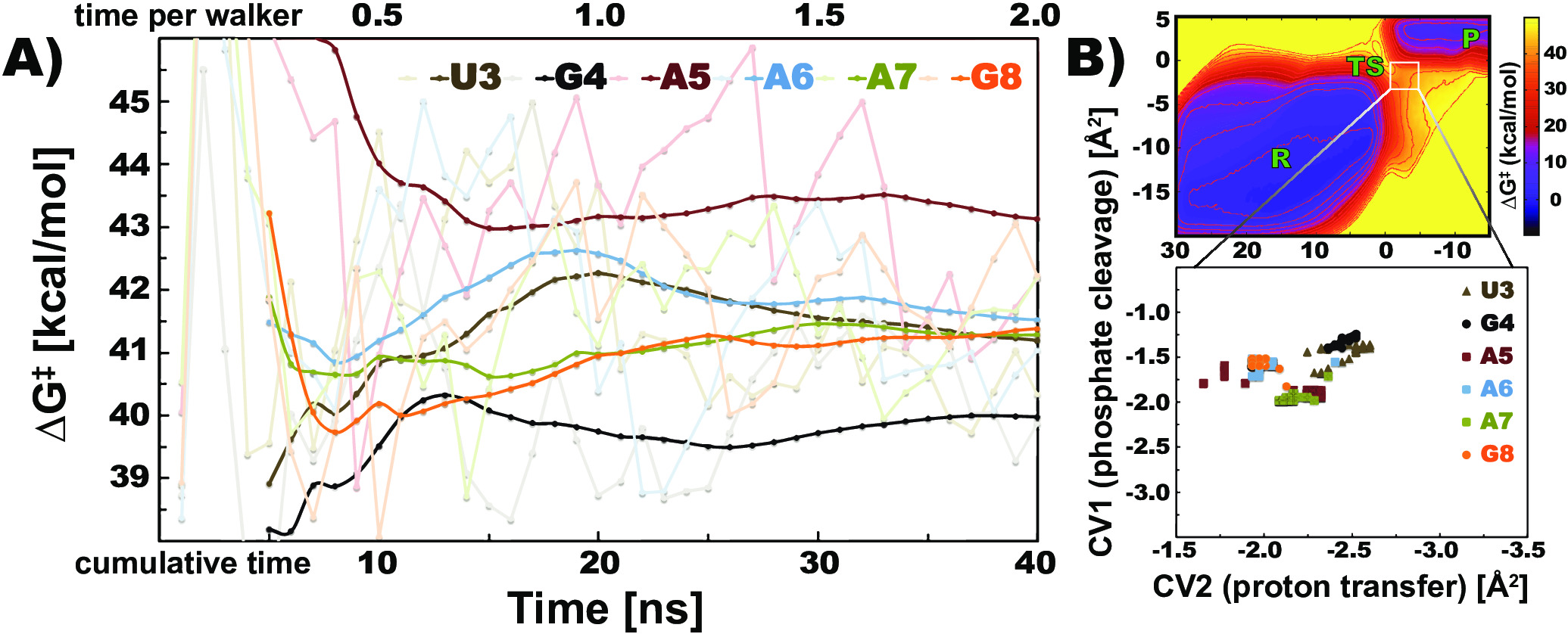}
\par\end{centering}
\caption{\label{fig2}Convergence of cleavage barriers and distribution of
TS positions of nucleotides from uGAAAg tetraloop. (A) Instantaneous
estimates of $\Delta G^{\ddagger}$ barrier (shaded lines) fluctuate,
whereas barriers calculated from time-averaged bias potentials (see
\emph{Methods}) are converged (bold lines). Each uGAAAg nucleotide
is displayed in a specific color with two (shaded and bold) tones.
Horizontal axis has two scales, showing timescale of single simulation
(one walker, upper scale) or cumulative time considering 20 concurrent
QM/MM-MetaD simulations (lower scale). The $\Delta G^{\ddagger}$
values were analyzed every 1 ns in total timescale. (B) Position of
TS states on FES is not changing significantly during QM/MM-MetaD
simulations. TS states of all tested nucleotides from uGAAAg tetraloop
are located within the same area with minimal changes of CV2 (describing
the proton transfer). Colors for each nucleotide and the interval
for analysis correspond with those on panel (A).}
\end{figure*}

We tested carefully the statistical convergency of computed phosphodiester
cleavage barriers. The analysis showed that $\Delta G^{\ddagger}$
barriers of 6 investigated nucleotides within uGAAAg tetraloop fluctuate
within a few kcal/mol over the time of the simulation (shaded lines
at Figure \ref{fig2}A), making it difficult to differentiate among
nucleotides. All those barriers were estimated by using the final
bias potential. In order to increase the accuracy of the method, we
then calculated $\Delta G^{\ddagger}$ from time-averaged bias potentials.
This latter approach gives a smoother convergence, which enables nucleotides
to be clearly distinguished. The resulting $\Delta G^{\ddagger}$
barriers of uGAAAg nucleotides are clearly converged after 40 ns of
cumulative simulated time (Figure \ref{fig2}A). Note that even initial
estimated averages ($\sim$7 ns of total simulated time) show clear
differences within $\Delta G^{\ddagger}$ among tested nucleotides.
The GC-duplex was used as a control simulation because the computed
$\Delta G^{\ddagger}$ barriers are expected to be identical for three
equivalent nucleotides. Our approach shows that the computed $\Delta\Delta G^{\ddagger}$
differences of equivalent G and C nucleotides are negligible after
40 ns, i.e., up to 0.4 and 0.5 kcal/mol, respectively (Figure S3A
in SI). We also analyzed the location of TS during different stages
of the simulations because the estimation of $\Delta G^{\ddagger}$
depends on the position of TS and R states on the FES. The TS positions
of all nucleotides within uGAAAg tetraloop are located within a small
region in the CV space (Figure \ref{fig2}B). The variance in positions
of R states were even smaller (data not shown). The same trend was
observed for the nucleotides within GC-duplex (Figure S3B in SI),
whereas differences in TS positions on FES were slightly larger for
nucleotides within cUUCGg tetraloop (Figure S4B in SI). 

\begin{figure*}
\begin{centering}
\includegraphics[width=0.6\textwidth]{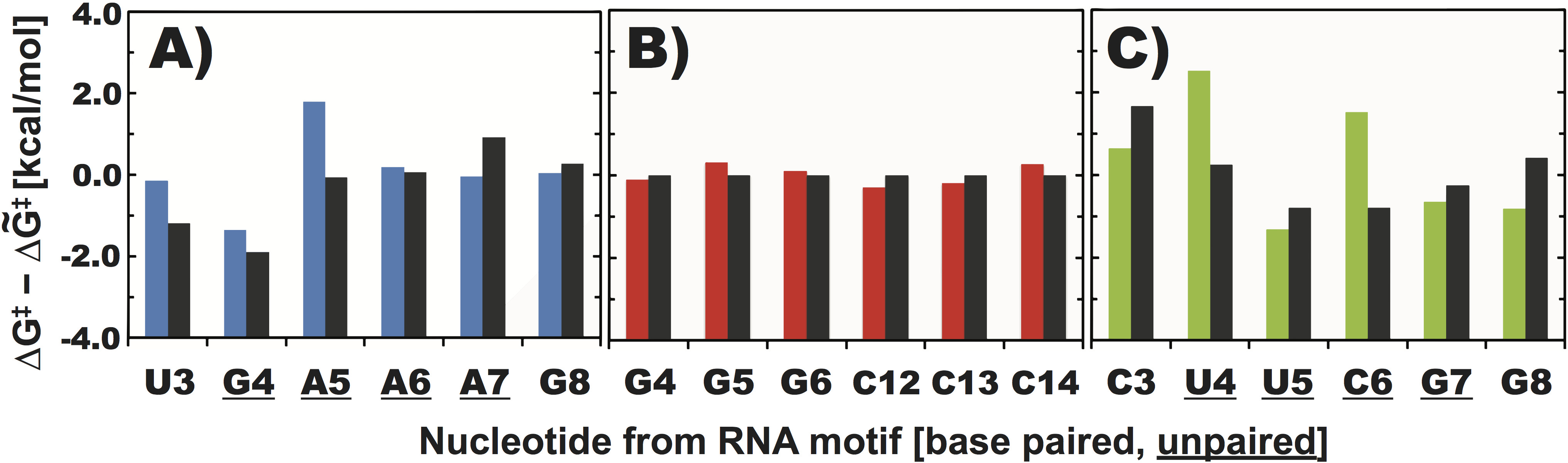}
\par\end{centering}
\caption{\label{fig3}Comparison between computed and experimentally derived
cleavage barriers for nucleotides from three different motifs. Correlation
between computed $\Delta G^{\ddagger}$ barriers (with specific color
for each motif) and barriers derived from experiments (black, pseudo-free
energies, see \emph{Methods} for details) for nucleotides from uGAAAg
tetraloop (A), GC-duplex (B), and cUUCGg tetraloop (C). Both computed
and experimentally derived $\Delta G^{\ddagger}$ barriers are displayed
as deviations from the corresponding median value. Note that in-line
probing data for a uniform GC-duplex are not available, but equivalent
G-C base pairs are expected to have similar $\Delta G^{\ddagger}$
barriers (deviations from the median are set to zero in the figure).
Unpaired nucleotides from tetraloops are underlined (see Figure S1
in SI for nucleotide labelling).}
\end{figure*}

The phosphodiester backbone cleavage $\Delta G^{\ddagger}$ barriers
are generally between 41 and 42 kcal/mol, but some nucleotides showed
intrinsic differences. G4 within uGAAAg tetraloop revealed the lowest
barrier among all the nucleotides (40 kcal/mol, Figure S2 in SI).
The following A5 showed a significantly higher barrier of 43.1 kcal/mol,
whereas the $\Delta G^{\ddagger}$ barriers for the remaining nucleotides
were comparable ($\sim$41.5 kcal/mol). All the nucleotides within
the GC-duplex revealed comparable barriers between 40.9 and 41.5 kcal/mol.
Among cUUCGg nucleotides, U5 showed lowest barriers (40.6 kcal/mol),
whereas U4 and C6 provided significantly higher barriers of 44.5 and
43.5 kcal/mol, respectively (Figures S2 and S4A in SI).

We then compared the computed $\Delta G^{\ddagger}$ barriers against
data from the in-line probing measurements. The experimental reactivities
(pseudo free-energies) were derived from available polyacrylamide
gel electrophoresis (PAGE) datasets, quantified and normalized separately
according to the scheme described in \emph{Materials and Methods}.
We observed good agreement between computed and experimental reactivities
for nucleotides within the uGAAAg tetraloop. Our computations overestimated
the $\Delta G^{\ddagger}$ barriers for U3 and A5 nucleotides, but
led the an overall correct trend (Figure \ref{fig3}A). However, a
similar comparison for the nucleotides within cUUCGg tetraloop revealed
some differences. U5, C6, and G7 were identified as reactive nucleotides
according to the experiment (\cite{Strauss_DAse-rib_NAR2012}), but
the computed $\Delta G^{\ddagger}$ barrier for C6 was significantly
higher, suggesting that the nucleotide is non-reactive (Figure \ref{fig3}C).
Note that the in-line probing data for a uniform GC-duplex are not
available, but that reactivity of paired residues is typically lower
than reactivity for unpaired residues (\cite{Soukup-Breaker_RNA1999,DAmare_c-di-GMP-riboswitch_NChemBiol2009,Strobel-glycine-riboswitch_RNA2011,Strauss_DAse-rib_NAR2012,Nelson_ydaO-rib_NatChemBiol2013,Hammond-ILP-SAM1riboswitch_ChemBiol2014,Breaker-riboswitches_MolCell2015}).

\section{Discussion}

In this paper, we used QM/MM-MetaD calculations to classify in-line
probing experiments characterized by the phosphodiester backbone cleavage
reaction. To this aim, we calculated the free-energy profiles modeling
the RNA backbone cleavage for 18 nucleotides within two RNA tetraloops
and a dsRNA. The computed $\triangle G^{\ddagger}$ obtained from
the QM/MM-MetaD calculation is expected to be related to the reactivity
of the particular nucleotide as observed in in-line probing experiments.
The aim of our study was not to predict absolute reactivities but
to explain differential reactivities observed among nucleotides within
the same or from different RNA motifs. To minimize the error in differential
estimates, we forced the system to explore a similar reaction pathway
for each nucleotide and prolonged simulations in the tens-of-ns timescale.
The approach provides converged and consistent results for all the
nucleotides. Results were then assessed by comparing $\triangle G^{\ddagger}$
barriers of identical nucleotides within a GC-duplex motif and by
analyzing reference in-line probing reactivities from PAGE gels.

We included a number of artificial restraints, which were required
in order to automatize the computational protocol, i.e., to allow
the straightforward comparison of various nucleotides within distinct
RNA motifs. Restraints improved stability of the simulations namely
by preventing spurious rupture of bonds and by excluding several unphysical
geometries detected during phosphodiester cleavage reaction. However,
all the backbone dihedrals, sugar puckers, glycosidic bonds, as well
as base pairing and stacking were left free to rearrange. We actually
observed significant dynamics during our QM/MM-MetaD simulations\emph{.}
This is important since the essence of in-line probing experiments
is to quantify the effects of RNA structural fluctuations on phosphodiester
cleavage rate. The introduction of restraints is nevertheless expected
to affect the computed barriers, and thus the respective cleavage
rates, in two specific ways. Firstly, restraints forced the cleavage
reaction to proceed through the designed reaction pathway that is
similar for all nucleotides and likely different from the validated
in-line attack reaction pathway. The possible contributions of other
reaction pathways, that could be different between one nucleotide
to the other, were thus ignored. As a result, the reactivity trends
estimated as differences of $\triangle G^{\ddagger}$ barriers could
in principle be compressed. Secondly, all the computed $\triangle G^{\ddagger}$
barriers are expected to be overestimated by excluding the scenario
where the proton from 2'-OH group is transferred through nbO atoms.
Considering the typical timescale of in-line probing experiments (hours/days),
the $\triangle G^{\ddagger}$ barriers derived from the estimated
rate constants using the Eyring equation are expected to be in range
from $\sim$26 to $\sim$32 kcal/mol (\cite{Soukup-Breaker_RNA1999}),
i.e., by $\sim$7 to $\sim$13 kcal/mol lower than the herein reported
$\triangle G^{\ddagger}$ barriers. Both those issues are very difficult
to tackle since they depend on intrinsic deficiencies of the DFTB3
parameterizations. Despite the fact that DFTB and other SE methods
improved significantly during last decade (\cite{York_puckers_JCTC2014,Christensen_Kubar_ChemRev2016}),
their general application towards chemical reactions remains challenging.
In particular, recent studies carefully assessed the performance and
revealed limitations of DFTB methods in description of phosphoranes
and phosphoryl transfers (\cite{Mlynsky_JCTC2014,Gaus_3obset-jctc2014,Huang-York_JCC2015,Cui_qmmm-SE-methods_MolSim2016}).
In this study, we still opted for the usage of DFTB3 because we did
not aim for an accurate description of states along phosphodiester
cleavage reaction. We rather focused on relative differences of $\triangle G^{\ddagger}$
barriers among different nucleotides, forcing the reaction to proceed
through the same pathway for all the analyzed nucleotides. We expect
such an approach to be more robust and less sensitive to the applied
QM method. QM/MM-MetaD simulations at least on the several-ns timescale
are required to converge these FES computations to a level allowing
for the discrimination of reactivity patterns, ruling out more accurate
QM methods such as DFT or \emph{ab initio}.

\begin{figure}
\begin{centering}
\includegraphics[width=0.6\columnwidth]{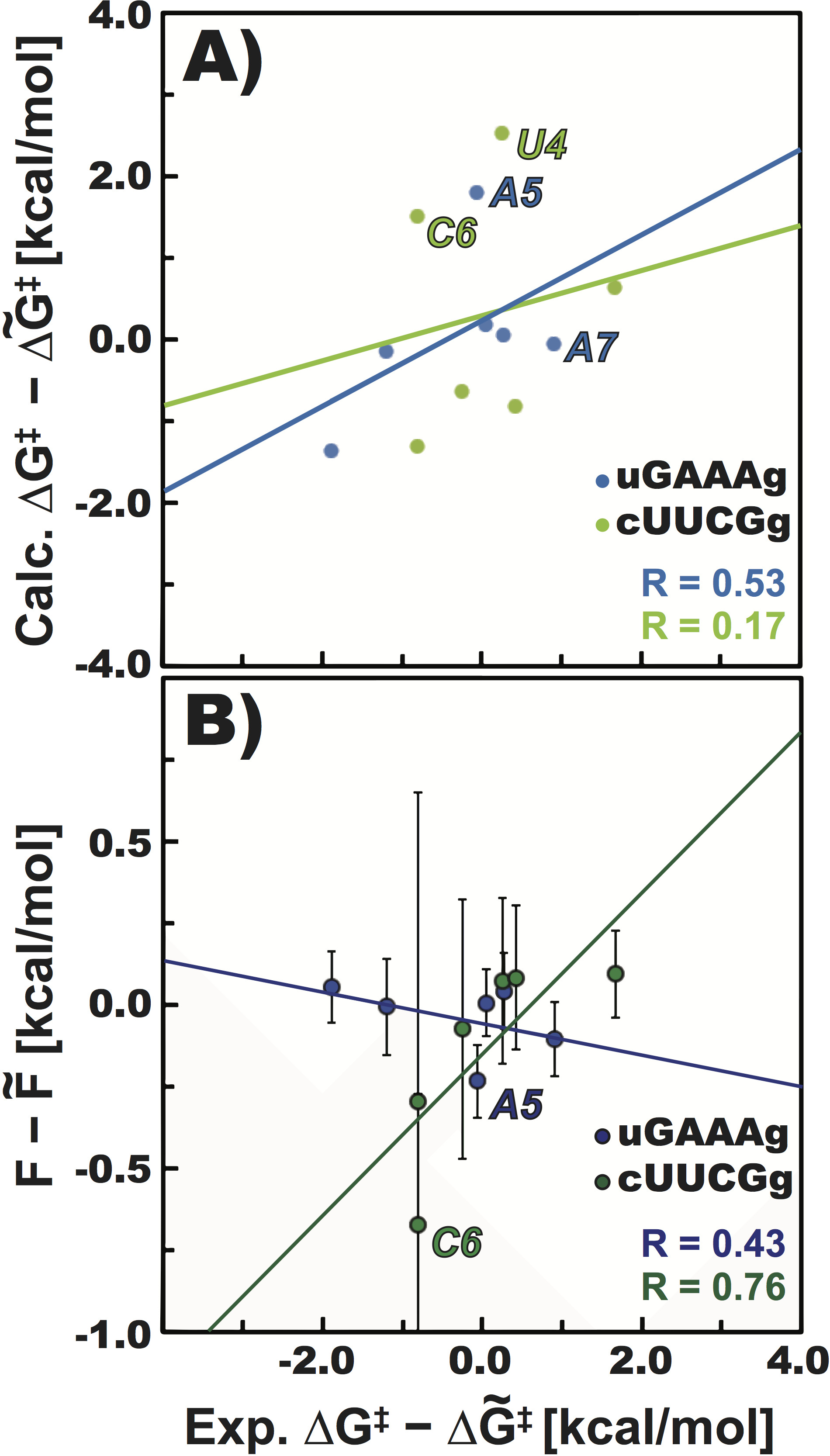}
\par\end{centering}
\caption{\label{fig4}Direct correlation between theory and experiments for
nucleotides within two tetraloops. (A) The correlation between calculated
$\triangle G^{\ddagger}$ and barriers derived from experiments for
nucleotides within uGAAAg (blue) and cUUCGg (green) tetraloops. Both
computed and experimental barriers are displayed as deviations from
the corresponding median value. (B) A similar correlation plot obtained
using the logarithm of in-line fitness $F=-RT\ln(\mbox{Fitness})$,
(\cite{Soukup-Breaker_RNA1999}), instead of the here calculated $\triangle G^{\ddagger}$.
We displayed averaged values of fitness from the set of X-ray structures
(considering 41 uGAAAg and 754 cUUCGg tetraloops, respectively) from
PDB database with bars showing their standard deviations. F values
are displayed the same way as $\triangle G^{\ddagger}$ barriers using
medians. Points corresponding to specific residues that are worsening
the correlation (outliers) are labelled.}
\end{figure}

Here, we ranked different RNA nucleotides, i.e., base paired/unpaired,
purine/pyrimidine, from duplex and tetraloops by their tendency to
undergo phosphodiester backbone cleavage. Nucleotide reactivities
reported by in-line probing experiments were used as a reference.
A number of experimental datasets for specific motifs from different
RNA systems are available (Figure S5 in SI, \cite{Soukup-Breaker_RNA1999,DAmare_c-di-GMP-riboswitch_NChemBiol2009,Strobel-glycine-riboswitch_RNA2011,Strauss_DAse-rib_NAR2012,Breaker-riboswitches_MolCell2015}),
but their quantitative estimation is often affected by unclear signals
(\cite{Strobel-glycine-riboswitch_RNA2011,Hammond-ILP-SAM1riboswitch_ChemBiol2014,Breaker-riboswitches_MolCell2015})
and/or participations in a tertiary interaction within complex RNA
structure (\cite{Soukup-Breaker_RNA1999,DAmare_c-di-GMP-riboswitch_NChemBiol2009,Strobel-glycine-riboswitch_RNA2011}).
Hence, we used a single specific experiment providing distinct signals
for all nucleotides as a reference of each of the tetraloop motifs
(\cite{Strauss_DAse-rib_NAR2012,Nelson_ydaO-rib_NatChemBiol2013}).
We observed that nucleotides within cUUCGg tetraloop revealed some
differences between predicted and calculated reactivities (Figure
\ref{fig3}C). Calculated $\triangle G^{\ddagger}$ barriers of U4
and C6 nucleotides are significantly overestimated, resulting in a
poor correlation between theory and experiment (R=0.17, Figure \ref{fig4}A).
We notice that (i) the FES profiles for cUUCGg nucleotides are statistically
converged (Figure S4A in SI), (ii) experimental reactivities for nucleotides
within cUUCGg motif reveal similar trends among different systems
(Figure S5B in SI), and (iii) the procedure of quantification of experimental
reactivities, i.e., using digitalized images (see \emph{Materials
and Methods}) provide almost identical profile against the raw count
data (a comparison for uGAAAg tetraloop is reported, Figure S6 in
SI). Thus, the poor correlation between computed $\triangle G^{\ddagger}$
barriers and experimental reactivities for cUUCGg suggests possible
limitations of our approach. We carefully inspected the structures
along the cleavage reaction and found that the reactive 2'-OH of U5
and, especially, U4 established H-bonds with other RNA groups outside
the QM region (described by the empirical force field, Figure S7 in
SI). Such interactions could result in over-stabilization of R states,
leading to the overestimation of the computed $\triangle G^{\ddagger}$
barriers by the current approach. Furthermore, cleavage site of C6
favored rare conformations with high in-line attack angle. Such geometry
is not favorable for the mechanism enforced herein and would require
to explore the in-line attack reaction pathway, where nbO atoms (and/or
external RNA groups, water molecules) are involved in the proton transfer.
This was not possible due to deficiencies within DFTB3 parameterization
(see SI for details). One possible way to improve the results for
cUUCGg would require the number of atoms within QM region (described
by DFTB3) to be increased. However, we identified that RNA groups
forming those interactions are typically belonging to nucleotides
located further away along the sugar-phosphate backbone, making the
calculation unfeasible. 

On the other hand, the agreement between theory and experiment is
satisfactory for both GC-duplex and uGAAAg tetraloop. In the former
case, nucleotides revealed very similar $\triangle G^{\ddagger}$
barriers, which is expected for three consecutive equivalent G and
C within dsRNA. The fact that each of these barriers was obtained
with a totally independent simulation further confirms the low statistical
error and hence the reproducibility of our computational approach.
The possible differences between purine and pyrimidine nucleotides
were negligible for this motif (within the error of our approach,
Figure S3A in SI). However, we observed that those $\triangle G^{\ddagger}$
barriers are comparable with cleavage barriers of several unpaired
residues within tetraloop motifs. This is unexpected, since the reactivity
of paired nucleotides is generally lower (\cite{Reynolds_NAR1996,Welch_Biochemistry1997,Soukup-Breaker_RNA1999}).
We speculate that such behavior is caused by the number of restraints
used in our computations, although it could also be linked to the
approximations in the SE method used. In the uGAAAg tetraloop, QM/MM-MetaD
simulations can clearly predict that the unpaired G4 is significantly
more reactive than the other nucleotides. Simulations are long enough
to consider this difference statistically significant. Other nucleotides
have higher $\triangle G^{\ddagger}$ barriers, in agreement with
the lower reactivity observed in experiments (Figure \ref{fig3}A).
Plotting computational and experimental reactivities against each
other revealed that the two unpaired nucleotides (A5 and A7) are worsening
the correlation (R=0.53) due to slightly overestimated (A5) and underestimated
(A7) $\triangle G^{\ddagger}$ barriers (Figure \ref{fig4}A). It
is worth noting that the overall stability and flexibility of the
tetraloop motifs can be affected by the nature of the closing base
pair (\cite{Bevilacqua_tloops-stab_Biochemistry2002,Bevilacqua_tloops-stab_JACS2009}).
For this reason, we explicitly replaced the G-C base pair observed
in the crystal structure with the wobble G-U pair found in the sequence
used in the reference in-line probing experiments.

Our results can be also compared against predictions made using the
approach introduced by Soukup and Breaker, where a correlation between
geometrical parameter (in-line fitness) and cleavage rates was proposed
(\cite{Soukup-Breaker_RNA1999}). The fitness parameter combines the
in-line attack angle and the distance between O2' and the adjacent
phosphorus. Interestingly, the fluctuations of the in-line attack
angle were proposed as a proxy for the chemical reactivity of individual
nucleotides (\cite{Sanbonmatsu_Methods-in-Enzymology2015}). We searched
among high-resolution X-ray structures ($\leq$3.5 \AA) of RNA molecules
from the RCSB Protein Data Bank (PDB, \cite{PDB-database_NAR2000})
and used the baRNAba tool (\cite{bottaro_baRNAba_NAR2014}) to extract
representative uGAAAg tetraloops. The average fitness values for those
nucleotides are anti-correlated with the experimental reactivities
derived from gels (Figure \ref{fig4}B), showing that the in-line
fitness formula (\cite{Soukup-Breaker_RNA1999}) derived from static
X-ray structures cannot reproduce the experimental reactivity pattern
for this motif. This is not surprising, since the in-line fitness
was not designed for differentiating among random nucleotides, but
rather for the specific identification of highly reactive nucleotides
within catalytic centers of ribozymes (see Figure 8 in the original
paper, \cite{Soukup-Breaker_RNA1999}).

Inspection of the starting structure used for uGAAAg computations
(PDB ID 4QLM, \cite{Ren_ydaO-x-ray_NatChemBiol2014}) revealed accidental
\emph{syn} orientation of the unpaired A5 and A7, which surprisingly
improved the correlation between in-line fitness and experimental
reactivity (R=0.85, Figure S8A in SI). We recall that considering
the structures extracted from the PDB as well as solution experiments
(\cite{Pardi_GNRA-tetraloops-NMR_SCience1991,Pardi_GNRA-tetraloops-NMR_NAR1996,Bottaro-RNA-in-stop-motion_NAR2016}),
the \emph{syn} conformation is rare and unexpected for nucleotides
within uGAAAg tetraloop. Since we did not find any apparent crystallographic
contact that may invoke those reorientations, it appears likely that
the higher flexibility of unpaired bases affected the refinement and
resulted in poor electron densities for unpaired nucleotides located
away from the important (binding) centers of the \emph{ydaO} riboswitch.
\emph{Syn/anti} flips of A5, A6 and A7 nucleotides also occurred during
classical MD simulation used for the system equilibration. Remarkably,
all nucleotides revealed the correct \emph{anti} conformation after
100 ns of the simulation time, i.e., in the structure used for subsequent
QM/MM-MetaD calculations, indicating that the MM force field used
was able to recover the expected native structure. We notice that
the \emph{syn/anti} flips of all unpaired nucleotides from uGAAAg
tetraloop were also occasionally detected during QM/MM-MetaD simulations
(in timescale of tens to hundreds of ps), despite the fact that the
starting structure contained all bases in correct \emph{anti} conformation.
This may suggest that possible \emph{anti/syn} reorientation might
induce the phosphodiester backbone cleavage by enabling more favorable
ribose pucker state (C2'-endo) for the initial nucleophilic attack
and/or different conformations of the adjacent phosphate. Interestingly,
functional nucleotides within catalytic centers of RNAs are frequently
found in \emph{syn} conformation (\cite{Bevilcqua_syn-bases_RNA2011}).

In conclusion, we presented an approach to characterize the reactivity
in RNA motifs. We employed QM/MM calculations with semi-empirical
methods, in combination with multiple-walkers metadynamics, to compute
the free-energy barriers associated with phosphodiester backbone cleavage
in generic, non-catalytic nucleotides. The computational protocol
is fast and robust, though limited by the currently available parameters
for the DFTB3 method. Remarkably, our procedure was able to reproduce
and explain differential reactivities in a common RNA tetraloop (uGAAAg).
However, reactivities in another tetraloop (cUUCGg) were more difficult
to classify. Our results suggest that better DFTB3 parameters would
be required for appropriate modeling of phosphodiester cleavage reactions
of this system and our protocol may serve as a benchmark for the further
improvements of the semiempirical method. To the best of our knowledge,
this study represents the first computational approach for the interpretation
and classification of chemical probing experiments. The introduced
procedure could be applied to more complex RNA motifs, providing the
initial step for fast and cheap distinguishing among several experimentally
suggested RNA structures.

\section{Materials and Methods}

Initial structures of RNA motifs were taken from crystal structures,
i.e., PDB ID 4QLM (uGAAAg tetraloop, \cite{Ren_ydaO-x-ray_NatChemBiol2014}),
1QCU (dsRNA, \cite{Klosterman_GC-duplex-Xray_Biochemistry1999}) and
4JF2 (cUUCGg tetraloop, \cite{Liberman_preQ1_NatChemBiol2013}). Tetraloop
motifs contain 10 nucleotides and dsRNA duplex consists of 8 G-C basepairs
(see Figure S1 and Methods section in SI for structures and details).
The QM region included two ribose rings and two phosphates (Figure
\ref{fig1}A). We used the DFTB3 method (\cite{Gaus_DFTB3_JCTC2011}),
as implemented in GROMACS 5.0 (\cite{abraham2015gromacs,Kubar_JCC2015})
in combination with PLUMED (\cite{Tribello_Plumed2_CPC2014}). Recent
corrections (\cite{York_puckers_JCTC2014}) that improve the description
of ribose rings (sugar-puckers) were additionally applied using PLUMED.
Bases were described at the MM level by AMBER \emph{ff}14 (\cite{Cornell_parm94_JACS1995,Wang_parm99-JCC2000,Perez_bsc0_BiophysJ2007,Zgarbova_chiOL3_JCTC2011})
in order to handle all of them at the same level of theory. We explicitly
tested the performance of all available DFTB3 parameter sets, i.e.,
MIO (\cite{Gaus_DFTB3_JCTC2011}), 3OB (\cite{Gaus_3obset-jctc2014}),
and 3OB-OPhyd (\cite{Gaus_3obset-jctc2014}). After accurate validations
we opted for the MIO set and all results reported herein were calculated
by that setup. To avoid spurious reactions and unphysical geometries
we had to enforce specific reaction pathways with a number of artificial
restraints to disallow the rupture of bonds not involved in the cleavage
reaction. These restraints might penalize the reactive in-line attack
geometry and the enforced pathway is likely to be different from the
reaction monitored by in-line probing experiments (see SI for details).

Well-tempered metadynamics (MetaD, \cite{Laio_PNAS2002,Bussi_WTmeta_PRL2008})
under the multiple-walker algorithm (\cite{Raiteri_walkers_JCPB2006})
was used to accelerate the phosphodiester cleavage and to estimate
the associated FES. Two CVs were employed (Figure \ref{fig1}A): one
to describe the proton transfer and the other to describe the phosphodiester
cleavage. FES were computed either considering the final bias potential
or considering time-averages of the bias potential (\cite{Micheletti_average-bias_PhysRevLett2004})
and convergence was monitored during different stages of the simulation.
Further discussion and justification for the time-averaging procedure
can be found in the \emph{Methods} section of SI. The activation free-energy
($\Delta G^{\ddagger}$) barrier of the phosphodiester backbone cleavage
for the particular nucleotide was extracted from computed FES by localizing
the saddle point (TS) on the reaction coordinate, i.e., the most convenient
path (requiring the lowest energies) connecting two areas with minimal
energies on the FES, corresponding to R and P state geometries (Figure
\ref{fig1}).

Experimental reactivities for specific nucleotides within uGAAAg and
cUUCGg tetraloops were quantified by analyzing PAGE data. We took
digitalized images extracted from the original papers (\cite{Nelson_ydaO-rib_NatChemBiol2013,Strauss_DAse-rib_NAR2012})
and we integrated the color density present in the area of the image
corresponding to each nucleotide. Subsequently, pseudo free-energy
reactivities were derived using an approach similar to the one developed
for SHAPE experiments (\cite{Weeks_Methods2010}): $\Delta G_{Exp}=-m\ln s$,
where $s$ is the signal intensity from gels and $m=2.6$ kcal/mol.
Intensities were normalized by shifting the medians of experimental
reactivities to match the one of the calculated barriers for each
tetraloop motif. We notice that all the considered systems were studied
using identical settings and analysis procedures so as to allow for
an unbiased comparison.

\section{Supplemental Material}

Supplemental material is available for this article and contains detailed
Methods section, preliminary QM/MM-MetaD runs and additional figures
showing structures of investigated RNA motifs, FES's for all compared
nucleotides, convergence of barriers for GC-duplex and cUUCGg tetraloop,
quantitation of experimental cleavage patterns, sample structures
of R state for U4 and U5 nucleotides, additional correlation between
in-line fitness and nucleotide reactivity, and sample PLUMED input
file.

\section{Acknowledgement}

Tom\'{a}\v{s} Kuba\v{r} is acknowledged for providing early access
and support in using the DFTB3 implementation for GROMACS. Philip
C. Bevilacqua is acknowledged for carefully reading the manuscript
and providing useful suggestions. Ronald R. Breaker and James W. Nelson
are acknowledged for providing raw experimental datasets. Sandro Bottaro
and Alejandro Gil-Ley are also acknowledged for help with setting
up trajectory analysis and preparation of DFTB3 input in GROMACS,
respectively. The research leading to these results has received funding
from the European Research Council under the European Union's Seventh
Framework Programme (FP/ 2007-2013) / ERC Grant Agreement n. 306662,
S-RNA-S.

\bibliographystyle{apalike}

\end{document}